# Can We Observe Weak Anomalous Couplings of Heavy Quarks Through Three Jet Events? *

THOMAS G. RIZZO
*Stanford Linear Accelerator Center*
*Stanford University, Stanford, CA 94309*

**Abstract**

The rates and corresponding jet distributions for the decay $Z \to b\bar{b}g$ and the process $e^+e^- \to t\bar{t}g$ may be sensitive to anomalous dipole-like couplings of heavy quarks to the photon and $Z$. In the $b$-quark case, after updating our previous analysis on the constraints imposed by current experiments on $Zb\bar{b}$ anomalous couplings, we show that the variation of these couplings within their presently allowed ranges leads to rather minor modifications to the Standard Model expectations for $Z \to b\bar{b}g$ observables. In the $t$-quark case, significant deviations from the Standard Model predictions for $t\bar{t}g$ production at the Next Linear Collider are possible.

*Work supported by the Department of Energy, contract DE-AC03-76SF00515.

# 1 Introduction

The value of $R_b = \Gamma(Z \to b\bar{b})/\Gamma(Z \to hadrons)$ as measured at LEP remains[1] more than $2\sigma$ higher than that predicted by the Standard Model(SM) for top-quark masses in the range found by the CDF[2] and D0[3] collaborations, *i.e.*, $m_t = 180 \pm 12$ GeV. If confirmed by future measurements, this unexpected result may be the first, albeit indirect, signal for new physics beyond the SM. This situation has inspired a large amount of theoretical speculation on the structure of possible new physics scenarios which can explain this discrepancy[4] without disrupting the great successes of the SM elsewhere. It may be that the third generation fermions will soon begin to tell us just what this new physics might be.

In a recent paper[5], we analyzed the constraints on possible anomalous weak couplings of heavy fermions($c$, $\tau$, $b$) to the $Z$ imposed by the then-existing data. Specifically, we considered adding contributions to the conventional SM $f\bar{f}Z$ vertex due to the weak electric ($\tilde{\kappa}_f$) and/or magnetic ($\kappa_f$) anomalous moment type couplings[6], *i.e.*,

$$\mathcal{L} = \frac{g}{2c_w}\bar{f}\left[\gamma_\mu(v_f - a_f\gamma_5) + \frac{i}{2m_f}\sigma_{\mu\nu}q^\nu(\kappa_f - i\tilde{\kappa}_f\gamma_5)\right]fZ^\mu\,, \tag{1}$$

where $g$ is the standard weak coupling constant, $c_w = cos\theta_W$, $m_f$ is the fermion mass, and $q$ is the $Z$'s four-momentum. In the case of the top quark, such a possibility has been entertained by a number of authors[7]. Using the data from both LEP and SLD available at the completion of the 1994 summer conferences[8, 9], we found reasonably strict constraints on both $\tilde{\kappa}_f$ and $\kappa_f$ for $f = c, \tau$ but, in the $f = b$ case, we found that the data preferred $\tilde{\kappa}_b$ and/or $\kappa_b$ to be non-zero at the $\simeq 2\sigma$ level reflecting the deviation of $R_b$ from the expectations of the SM.

In this paper, after updating this analysis for $b$-quarks using the more recent data presented at Moriond95[10, 11], we will consider the feasibility of probing the $Zb\bar{b}$ vertex in the three-body $Z \to b\bar{b}g$ decay process[12]. We then extend this approach to the case of open top production at the Next Linear Collider(NLC). In order to perform the $b$-quark analysis, we need to know the currently allowed ranges of $\kappa_b$ and $\tilde{\kappa}_b$ which requires us to revise our previous study. Essentially, we expect that the effects of non-zero values for $\tilde{\kappa}_b$ and $\kappa_b$ are two-fold since both the overall value of the ratio $R = \Gamma(Z \to b\bar{b}g)/\Gamma(Z \to b\bar{b})$ as well as the corresponding decay distributions are modified. From the shift in the value of this ratio due to anomalous couplings, it would appear that the universality of the strong interactions is violated since the extracted value of $\alpha_s^b$ would be somewhat different from $\alpha_s^{udsc}$. Just how large these anticipated effects can be given the tight restrictions from LEP/SLC precision measurements is a subject of the present analysis. In the case of $e^+e^- \to t\bar{t}g$, the lack of any strong restrictions from existing data plays a crucial role. This means that *simultaneous* studies of $t\bar{t}$ and $t\bar{t}g$ final states, which are quite complementary, will be important at the NLC. Unlike the $b$-quark case, both anomalous $t\bar{t}Z$ as well at $t\bar{t}\gamma$ vertices are probed by high energy $e^+e^-$ collisions, and our analysis will compare the sensitivities to both types of anomalous couplings.

# 2 $Z \to b\bar{b}g$

As pointed out in Ref.5, if $\tilde{\kappa}_b$ or $\kappa_b$ were non-zero, a number of $Z$-pole observables would differ from the expectations of the SM. (A complete list of all such observables and their dependencies on $\tilde{\kappa}$ and $\kappa$ are given in detail in this reference.) In that analysis, we considered the following data as input: $R_b$ and $A_{FB}^b$(the forward-backward asymmetry), both measured at LEP, as well as $A_{pol}^b$(the polarized forward-backward asymmetry), which is measured by SLD. Fig.1 shows the results of the now updated version of our analysis for the ratios $R_b/R_b^{SM}$ and $A_b/A_b^{SM}$, where the latter quantity is the weighted combination of $A_{FB}^b/A_{FB}^b(SM)$ and $A^b_{pol}/A^b_{pol}(SM)$ (under the assumption that the electron's couplings are given by the SM expectations), when $\tilde{\kappa}_b$ and/or $\kappa_b$ are non-zero. In this analysis we have fixed $\alpha_s(M_Z) = 0.125$, $\alpha_{em}^{-1}(M_Z) = 128.896$[13], and the SM Higgs boson mass ($m_H$) to 300 GeV. A modified version of ZFITTER4.9[14] was used to obtain the predictions of the SM for these observables assuming $m_t = 170,\ 180$ or 190 GeV, providing us with the SM input in Fig.1. Allowing $\tilde{\kappa}_b$ and $\kappa_b$ to be non-zero, we can then perform a $\chi^2$ fit to determine the 95% CL region for these anomalous couplings, for fixed $m_t$, using the latest results from Moriond95[1, 10, 11]. Fig.2 shows the result of this updated analysis which we note is little influenced by variations of the input parameters other than $m_t$. As can be seen from this figure, the SM lies just outside the 95% CL region when $m_t = 180$ GeV and the data somewhat favors $\tilde{\kappa}_b$ and $\kappa_b$ non-zero with magnitudes of order $10^{-2}$. The SM lies on the boundary of the allowed region due to the $2\sigma$ discrepancy in the value of $R_b$. To clarify this issue, more data on all of the above observables is necessary and these will become available over the next two years. Unlike in the $b$-quark case, our updated analysis shows no shred of evidence of new physics in the corresponding $\chi^2$ fits for $c$ and $\tau$. In comparison to our published results which made use of the data set from the the 1994 summer conferences, the results from Moriond95 shrink the radii of the new 95% CL allowed regions by approximately 5% and 25% for $c$ and $\tau$, respectively.

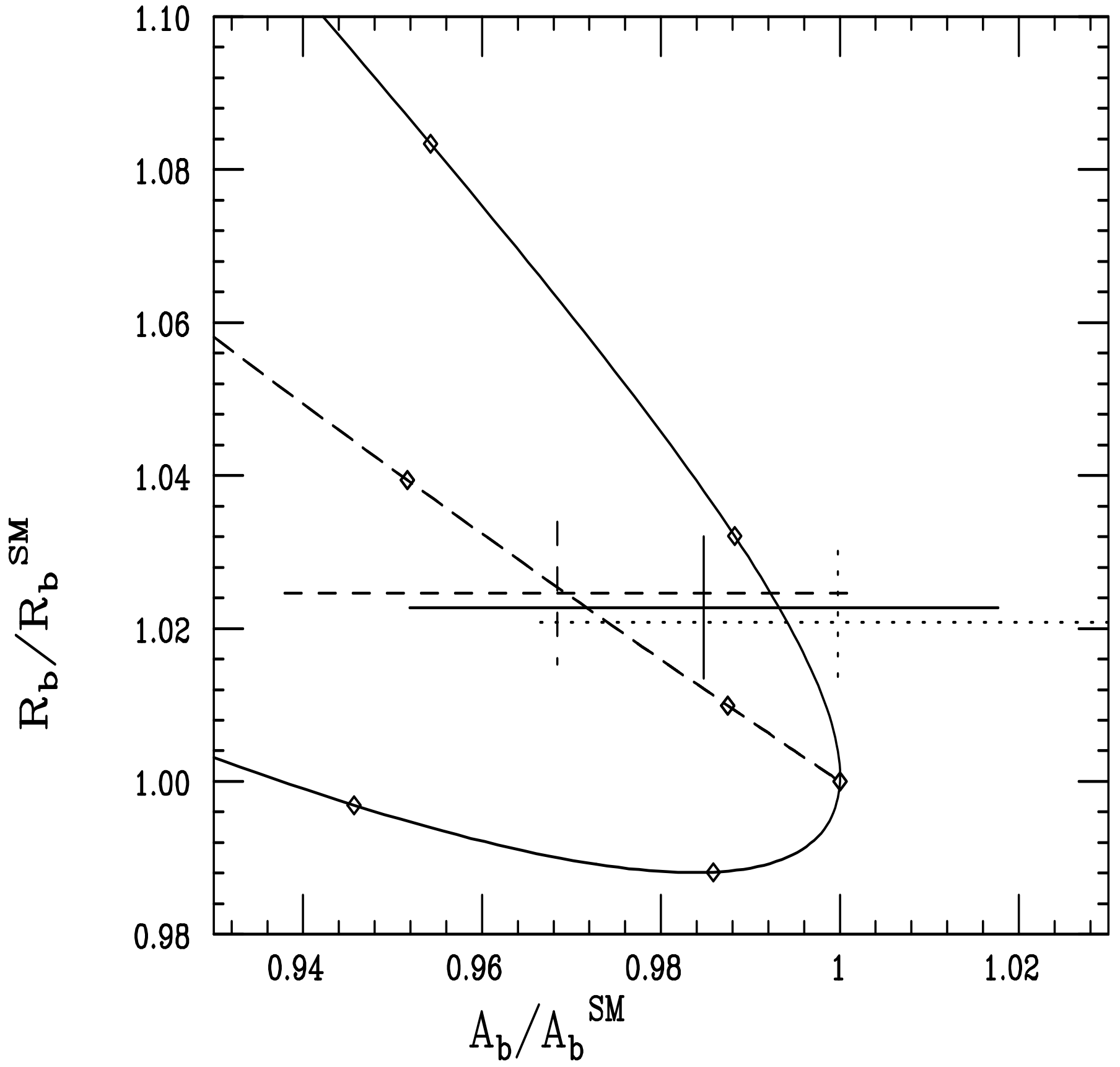

Figure 1: $R_b$ vs. $A_b$ compared with the predictions of the SM for $m_t$ = 170, 180, 190 GeV, corresponding to the dotted, solid, dashed data point, respectively. The upper(lower) solid curve is the prediction for non-zero negative(positive) values of $\kappa_b$ with the points in steps of 0.01. The dashed line represents the corresponding case of non-zero $\tilde{\kappa}_b$.

Is there any other way to probe the values for $\tilde{\kappa}_b$ and/or $\kappa_b$ in the above range other than through these traditional observables? One possibility, alluded to above, is to examine the the decay $Z \rightarrow b\bar{b}g$ as, *a priori*, we might expect that the modifications of the $Zb\bar{b}$ vertex may show up as deviations from SM expectations in both the rate and corresponding jet distributions. As we will see below, a leading order(LO) calculation is sufficient for our

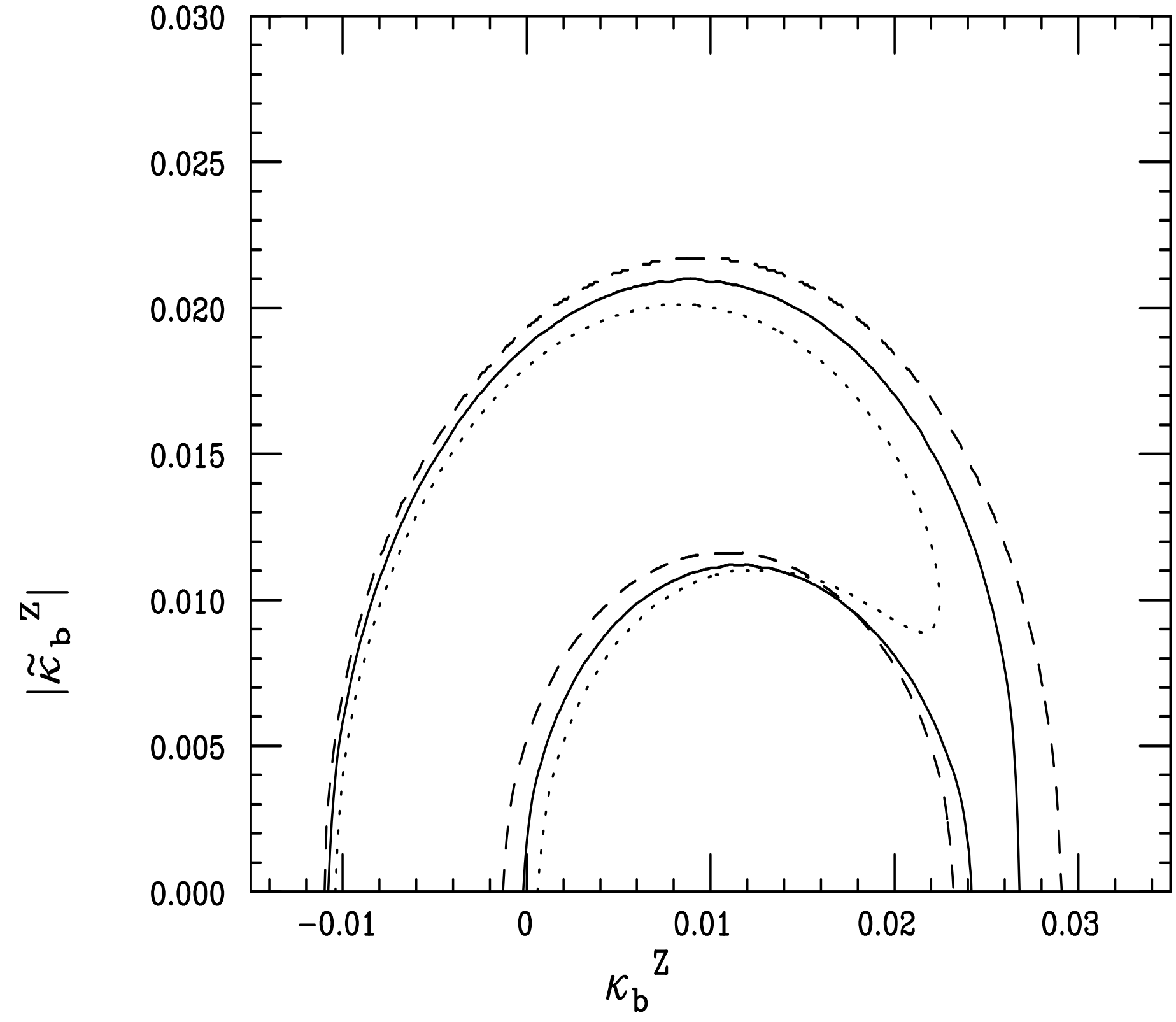

Figure 2: Regions in the $\kappa_b$-$\tilde{\kappa}_b$ plane allowed at the 95% CL by the Moriond95 data for $m_t$ = 170, 180, 190 GeV, corresponding to the inside of the dotted, solid, and dashed curves, respectively.

purposes. To this end we consider the double differential ratio

$$\frac{d^2R}{dx_1dx_2} = \frac{1}{\Gamma(Z\to b\bar{b})}\frac{d^2\Gamma(Z\to b\bar{b}g)}{dx_1dx_2}\,, \tag{2}$$

where, to leading order in $m_b^2/M_Z^2$, the individual components of this expression are given by (omitting an overall common normalization factor which cancels in taking the ratio)

$$\begin{aligned}
\Gamma(Z\to b\bar{b}) &\simeq (v_b^2+a_b^2)+\frac{r}{8}(\kappa_b^2+\tilde{\kappa}_b^2)+3v_b\kappa_b\,,\\
\frac{d^2\Gamma(Z\to b\bar{b}g)}{dx_1dx_2} &\simeq \frac{2\alpha_s(M_Z)}{3\pi}\Bigg[(v_b^2+a_b^2)\frac{x_1^2+x_2^2}{(1-x_1)(1-x_2)}+\frac{r}{8}(\kappa_b^2+\tilde{\kappa}_b^2)f_1\\
&\quad -\frac{1}{2}v_b\kappa_b f_2\Bigg]\,,
\end{aligned} \tag{3}$$

where $r = M_Z^2/m_b^2 \simeq 360$ and the functions $f_i = f_i(x_1,x_2)$, with $x_{1,2} = 2E_{b,\bar{b}}/M_Z$, are explicitly given by

$$\begin{aligned}
f_1 &= [2-2(z_1+z_2)+2z_1z_2+(z_1^2+z_2^2)-4z_1z_2(z_1+z_2)]/(z_1z_2)\,,\\
f_2 &= [-2(z_1^2+z_2^2)-4z_1z_2+12(z_1+z_2)-12]/(z_1z_2)\,,
\end{aligned} \tag{4}$$

where $z_i = 1-x_i$. Note that we recover the usual QCD result for massless quarks in the limit when the anomalous couplings vanish. Of course, for completeness all higher order terms in $1/r$ are kept in our analysis below, though their numerical influence on our results is quite minimal. The complete expressions used in this analysis are given in detail in the Appendix. An important feature of these equations is that $\tilde{\kappa}_b$ does not appear linearly since such a term would be a direct measure of $CP$-violation. To get at such terms we need to make use of the initial $e^-$ momentum or polarization direction or the $b$-quark decay products to form asymmetries. From the above equations, we see that it is the rather large value of $r$ that provides the enhanced sensitivity to the $b$-quark anomalous couplings. To obtain the $Z \to b\bar{b}g$ width as well as the various distributions the above double differential must be integrated over various weighting factors; these integrals are evaluated by introducing a cut on the invariant mass of any pair of jets. This procedure is not unique when finite quark masses appear in the final state, but we have chosen to use for convenience the definition $2p_i \cdot p_j \geq y_{cut}s$, $(i \neq j,\ i,j = 1-3)$, where $p_i$ is one of the three jet four-momenta and $s = M_Z^2$. Our results will of course depend somewhat on the value chosen for the $y_{cut}$ parameter.

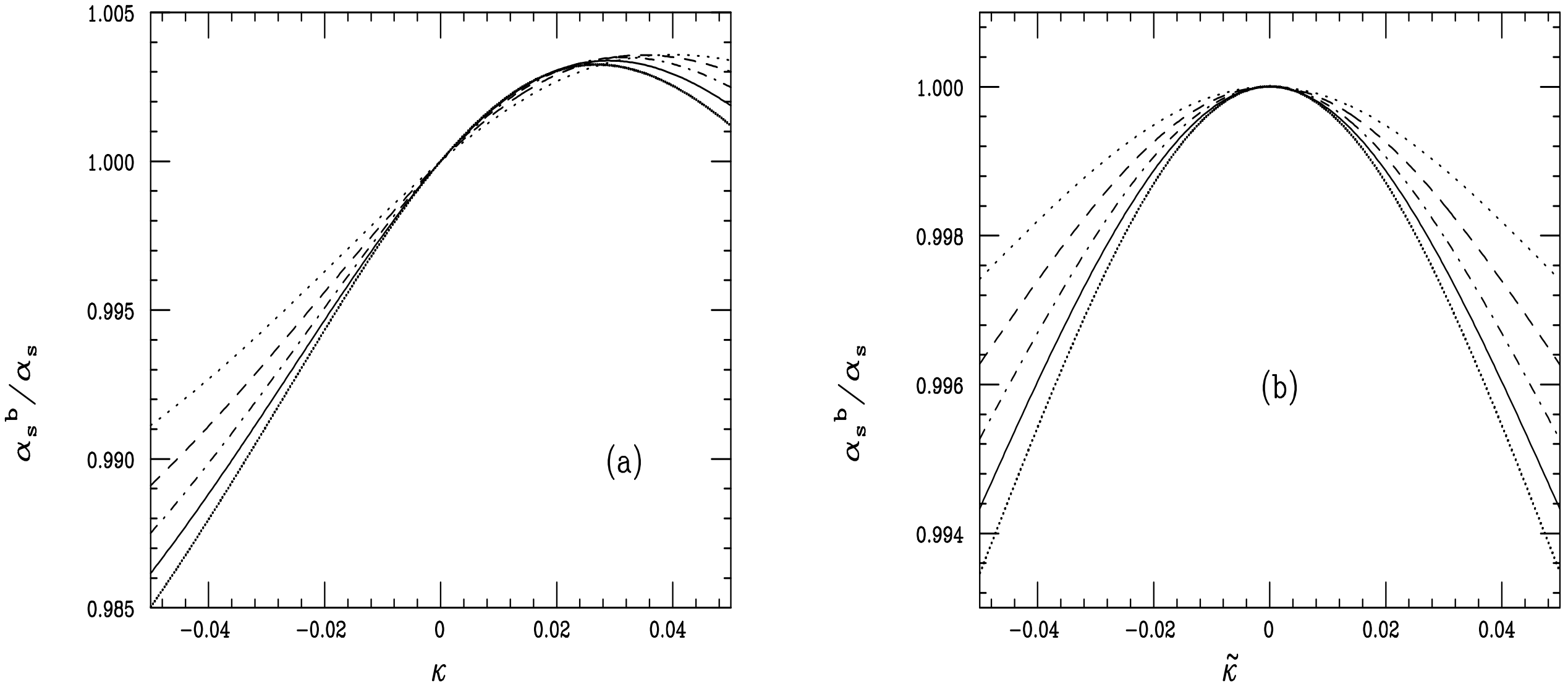

Figure 3: Values of $\alpha_s^b/\alpha_s^{udsc}$ due to non-zero (a) $\kappa_b$ or (b) $\tilde{\kappa}_b$ for $y_{cut}$ values from 0.01 to 0.05 in steps of 0.01 from top to bottom on the left side of the figures.

We first examine the $Z \to b\bar{b}g$ three jet rate. In order to directly compare with the SM, we will scale our results with non-zero $\kappa_b$ and $\tilde{\kappa}_b$ to the SM predictions for the same value of $y_{cut}$. One could interpret this ratio (in LO) as a measure of any apparent shift in the value of $\alpha_s$ for $b$-quarks in comparison to that for the lighter flavors, *i.e.*, as a test for violations of the flavor-independence of QCD. Figs. 3a and 3b show the individual $\kappa_b$ and $\tilde{\kappa}_b$ dependence of the $Z \to b\bar{b}g$ three jet rate for different values of $y_{cut}$ which we display as a shift in the value of $\alpha_s^b$ in comparison to the expectations of universality. The shift in the value of $\alpha_s^b$ is only at the percent level in either case. Note that since there is no linear term in $\tilde{\kappa}_b$, our results are an even function of $\tilde{\kappa}_b$ whereas the term linear in $\kappa_b$ remains quite important. Next, we scan the 95% CL allowed regions in the $\kappa_b$-$\tilde{\kappa}_b$ plane for $m_t = 170,\ 180$ or 190 GeV, shown in Fig.2, and ask how large a deviation from universality is allowed by the present electroweak data. We find that the rather restricted ranges of $\kappa_b$ and $\tilde{\kappa}_b$ do not allow for large violation in universality due to anomalous couplings. In particular, for $y_{cut} = 0.05$, we find that $0.997 \leq \alpha_s^b/\alpha_s^{udsc} \leq 1.004$ within this 95% CL region; essentially identical results are obtained for other values of $y_{cut}$. This implies that these apparent violations of universality are far smaller than what can be probed by current experiment. This is a direct result of the rather strong demands placed upon the anomalous couplings by the precision electroweak data. Present experimental analyses by SLD[15], ALEPH[16] and OPAL[17] find that $0.898 \leq \alpha_s^b/\alpha_s^{udsc} \leq 1.154$, $0.967 \leq \alpha_s^b/\alpha_s^{udsc} \leq 1.047$, and $0.969 \leq \alpha_s^b/\alpha_s^{udsc} \leq 1.073$, respectively, at the 95% CL. Naively combining these measurements in quadrature leads to $\alpha_s^b/\alpha_s^{udsc} = 1.013 \pm 0.028$ at 95% CL. We thus see that that the size of the deviations due to the presence of anomalous couplings is far below the present sensitivities (by about an order of magnitude) of these three experimental analyses, but may become visible in future data sets with significantly larger statistics and with greatly reduced systematic uncertainties.

Perhaps the various three jet distributions show a greater sensitivity to the existence of anomalous couplings than does the overall rate. To this end, we first consider the separate $x_{1-3}$ distributions where we now order $x_3 \leq x_2 \leq x_1$. Figs.4a-c show these three distributions for the two extreme non-zero values of $\kappa_b$ which are allowed at the 95% CL when $\tilde{\kappa}_b = 0$, *i.e.*, $\kappa_b = 0.027$ and $\kappa_b = -0.011$. Here, we make a direct comparison to the SM expectations, neglecting for simplicity the non-leading terms in $1/r$, which is numerically sufficient for our purposes. Except for slight differences in shape and normalization, these distributions do not significantly deviate from the SM predictions. Thus, it would appear that they are not too helpful in extracting information on anomalous couplings. To be complete, we also show in Fig.4d the distribution of the Ellis-Karliner angle[18] for the SM as well as the two extreme values of $\kappa_b$ above. As in the case of the $x_i$ distributions, we see that there is very little departure from the expectations of QCD when anomalous couplings are present.

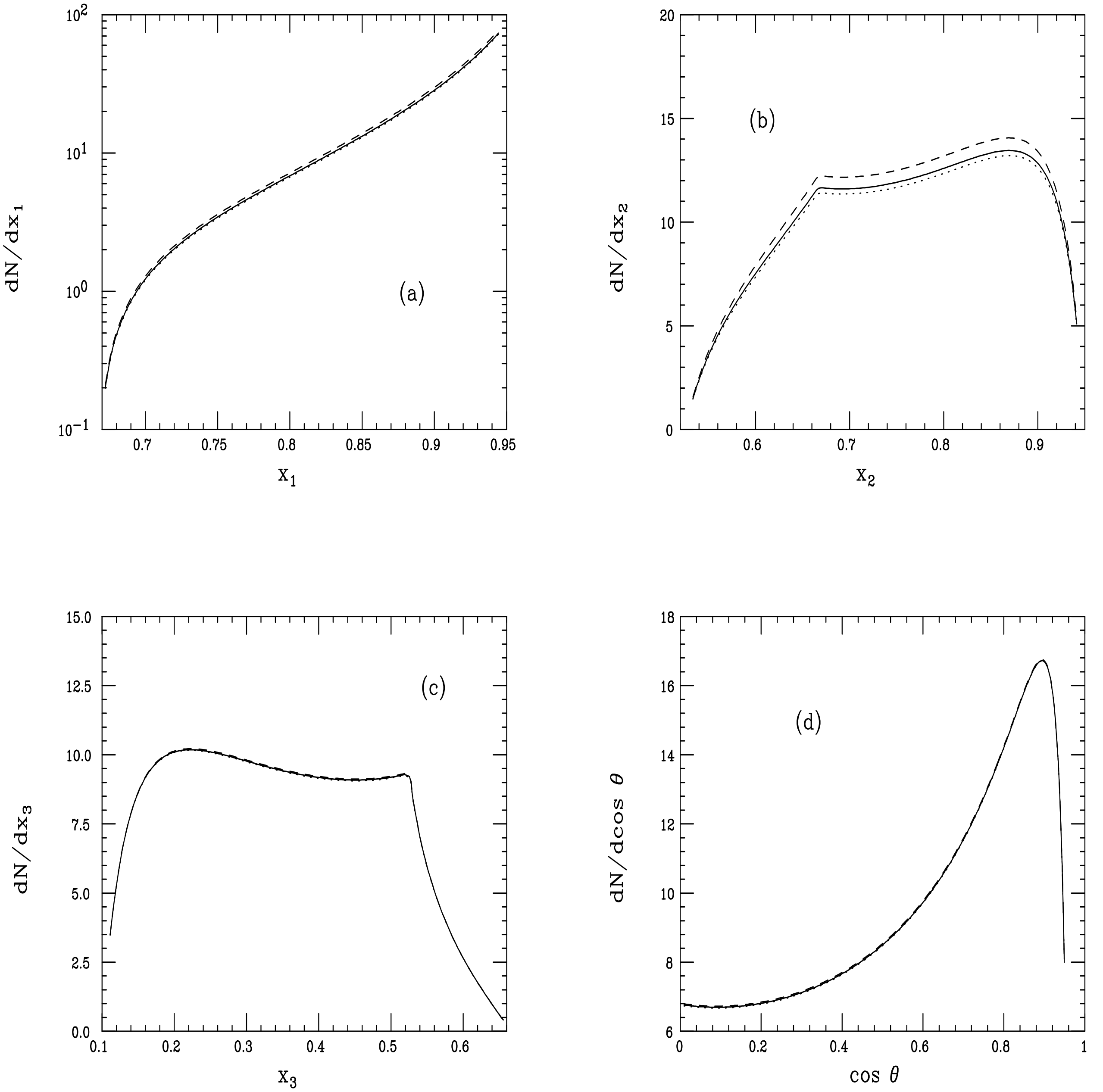

Figure 4: $x_i$ and Ellis-Karliner angle distributions for the SM(solid) as well as for $\kappa_b$ =0.027(dashed) and -0.011(dotted) with $\tilde{\kappa}_b = 0$. $y_{cut} = 0.05$ has been assumed.

As a last possibility we consider the gluon energy distribution itself in the case where the $b$ and $\bar{b}$ jets are tagged. While we do not anticipate *a priori* that this distribution is more sensitive that those above to the presence of anomalous couplings, we include it for completeness. (As we will see below, the gluon energy spectrum *will* yield important constraints in the top quark case.) Fig.5 confirms our expectations as it shows that the gluon energy spectrum as a function of $z = 2E_g/\sqrt{s}$ has little sensitivity to the existence of potential $Zb\bar{b}$ anomalous couplings.

We thus conclude in the case of $b$-quarks that the already existing high precision data does not allow for observable effects of anomalous couplings in $Z \to b\bar{b}g$ events. Even though these results are somewhat disappointing, one must continue to search for anomalous $b$-quark couplings in every possible manner.

## 3 $e^+e^- \to t\bar{t}g$

The situation for top is quite different than that for $b$'s as we are no longer sitting on the $Z$ pole and both $\gamma$ and $Z$ anomalous couplings may be present simultaneously. To obtain the distributions for this case we first define the coupling combinations

$$
\begin{aligned}
A_v &= \sum_{ij} (v_i v_j + a_i a_j)_e (v_i v_j)_t P_{ij} \,, \\
A_a &= \sum_{ij} (v_i v_j + a_i a_j)_e (a_i a_j)_t P_{ij} \,,
\end{aligned}
$$

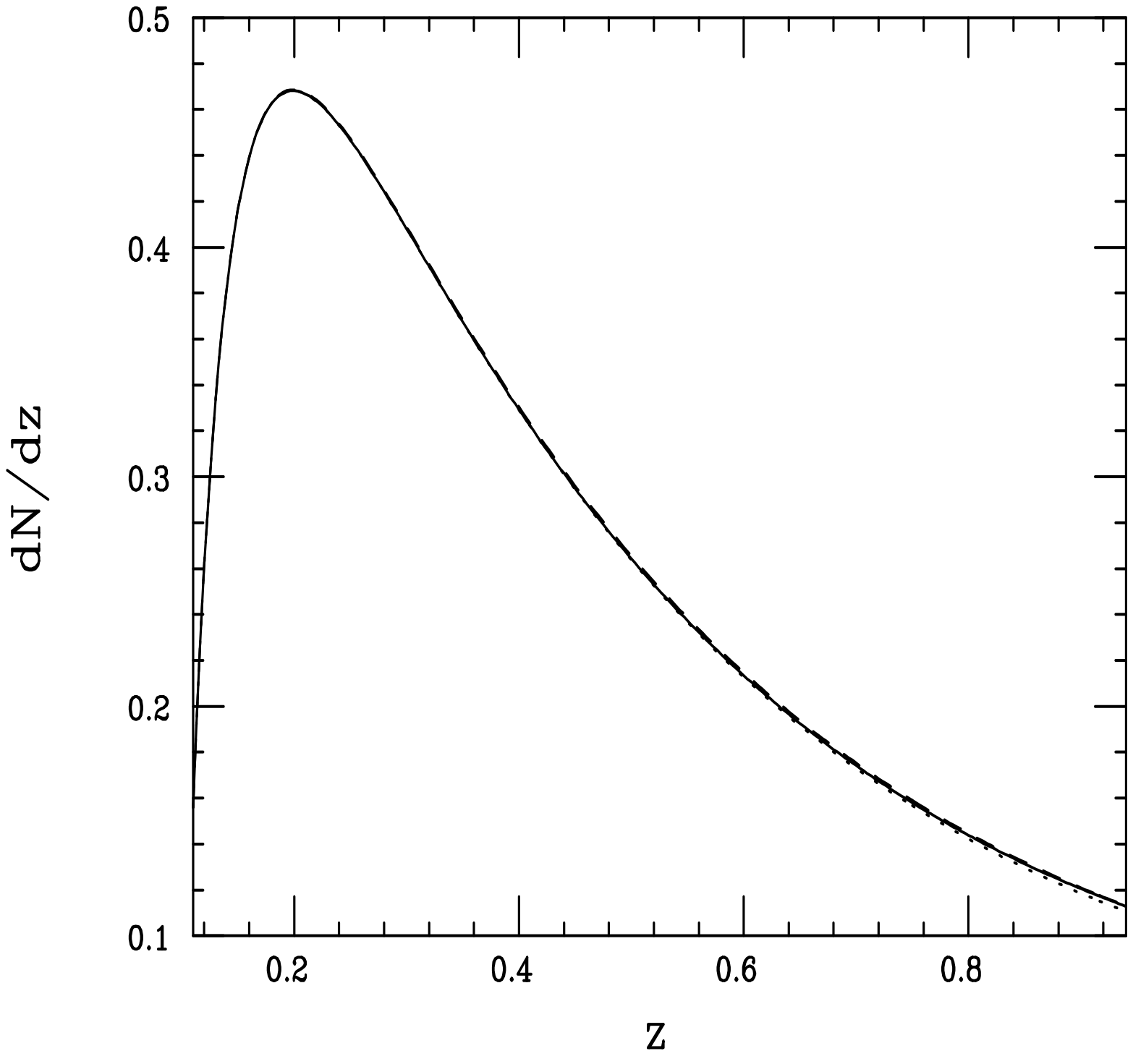

Figure 5: Gluon jet energy distribution in the case of tagged $b$-quarks for the SM(solid) as well as for $\kappa_b$ =0.027(dashed) and -0.011(dotted) with $\tilde{\kappa}_b = 0$. $y_{cut} = 0.05$ has been assumed and $z$ is the scaled gluon energy as defined in the text.

$$
\begin{aligned}
A_\kappa &= \sum_{ij} (v_i v_j + a_i a_j)_e (\kappa_i \kappa_j)_t P_{ij} \,, \\
A_{\tilde\kappa} &= \sum_{ij} (v_i v_j + a_i a_j)_e (\tilde\kappa_i \tilde\kappa_j)_t P_{ij} \,, \\
A_m &= \sum_{ij} \frac{1}{2} (v_i v_j + a_i a_j)_e (v_i \kappa_j + v_j \kappa_i)_t P_{ij} \,, \\
P_{ij} &= s^2 \frac{[(s - M_i^2)(s - M_j^2) + (\Gamma M)_i (\Gamma M)_j]}{[(s - M_i^2)^2 + (\Gamma M)_i^2][(s - M_j^2)^2 + (\Gamma M)_j^2]} \,, \qquad (5)
\end{aligned}
$$

where we sum over the contributions of both the photon and $Z$. $i = 1, 2$ labels the photon and $Z$ couplings respectively and thus $M_1 = \Gamma_1 = 0$ while $M_2 = 91.1887$ GeV and $\Gamma_2 = 2.4971$ GeV[10]. In analogy with the $b$-quark case we can write

$$
\frac{d^2 R}{dx_1 dx_2} = \frac{1}{\sigma(e^+e^- \to t\bar t)} \frac{d^2\sigma(e^+e^- \to t\bar t g)}{dx_1 dx_2} \,, \qquad (6)
$$

where

$$
\begin{aligned}
\sigma(e^+e^- \to t\bar t) \;&\simeq\; \frac{3\beta}{4} \Big[ (A_v + A_a)(1 + \beta^2/3) + (A_v - A_a)(1 - \beta^2) \\
&+ \frac{r}{4}(1 - \beta^2/3)(A_\kappa + A_{\tilde\kappa}) + A_\kappa - A_{\tilde\kappa} + 4A_m \Big] \,, \qquad (7)
\end{aligned}
$$

with $\beta^2 = 1 - 4m_t^2/s$ and $r$ is now given by $r = s/m_t^2$. Note that $r$ is no longer a large number. For the three body process we obtain

$$
\begin{aligned}
\frac{d^2\sigma(e^+e^- \to t\bar t g)}{dx_1 dx_2} \;&\simeq\; \frac{2\alpha_s(\sqrt{s})}{3\pi} \Big[ A_v f_{0v} + A_a f_{0a} + \frac{r}{8}(A_\kappa f_{1\kappa} + A_{\tilde\kappa} f_{1\tilde\kappa}) \\
&- \frac{1}{2} A_m f_2 \Big] \,, \qquad (8)
\end{aligned}
$$

with the $f$'s being the same kinematic functions given above and in the Appendix. In our numerical analysis below we will assume $\alpha_s = 0.10$ and neglect the possibility of initial beam polarization.

Since the top decays before it hadronizes, *i.e.*, $\Gamma_t$=1.57 GeV when $m_t$ =180 GeV[19], a true 3 body final state does not arise in $t\bar{t}g$ production. Therefore we cannot simply take our previous $b\bar{b}g$ jet analysis and apply it to top directly. For almost all observables of interest we must look for new physics in the distributions of the decay products of the top, *i.e.*, the $W$ and $b$. However, the gluon energy spectrum associated with $t\bar{t}g$ production *can* be used as a probe of anomalous couplings provided some care is used. The finite top width has several implications in addition to the consideration of the top decay products, including the fact that $\Gamma_t \neq 0$ acts as an infra-red regulator, just as $m_t \neq 0$ prevents collinear singularities. This softening of the spectrum near $z = 2E_g/\sqrt{s} = 0$ can be accounted for quite accurately by scaling all of the functions $f_i$ by a common factor of

$$\mathcal{F} = \frac{z_1^4 z_2^4}{(z_1^2+\delta^2)^2(z_2^2+\delta^2)^2}\,, \tag{9}$$

where $z_i$ is as defined after Eq.(4) and $\delta = m_t\Gamma_t/s \simeq 10^{-3}$ for a 500 GeV collider. Fig.6 shows the influence of finite $\Gamma_t$ in the SM for small values of $z$. Above $z \simeq 0.08-0.10$, corresponding to $E_g = 20-25$ GeV, the effect of the finite top width on this distribution becomes unobservable. This means that the emission of very hard gluons by top *before* it can decay are not very much influenced by the decay itself (*e.g.*, gluons that are emitted from the final state $b$-quarks) as long as $E_g \gg \Gamma_t$, a condition we will always impose below by demanding large values of $z$ in our analysis. If we want to look at distributions other than those associated with the gluon we must take the full top decay sequence into account.

As far as anomalous couplings are concerned, Fig.6 shows that all of the sensitivity to non-zero values of $\kappa_t^\gamma$, $\kappa_t^Z$ (and correspondingly $\tilde{\kappa}_t^\gamma$, $\tilde{\kappa}_t^Z$) occurs in the large $z \geq 0.10-0.15$ region of the spectrum, with the lower end showing only the universal SM effects. This low part of the $z$ spectrum is also useful, however, in that it can set the overall normalization. For simplicity, let us ignore $\tilde{\kappa}_t^\gamma$, and $\tilde{\kappa}_t^Z$ for now and concentrate on the magnetic weak dipole moments. In that case, Fig.6 shows that $\kappa_t^\gamma$ and $\kappa_t^Z$ with magnitudes of order 0.1 may be cleanly visible. We consider two possible approaches. First, one can just count the number of events above a given minimum value of $z$, $z_{cut}$, and compare with the SM. Second, one can bin the events above the cut and perform a fit to the spectrum to extract the anomalous couplings. The second possibility is far more sensitive as we shall see below. The results of the first procedure are presented in Figs.7a and 7b, where we display the normalized integrated $t\bar{t}g$ rate for $z > 0.2$ as a function of either of the two anomalous weak magnetic dipole moments. In the photon case we see that even with these highly optimistic assumptions only a small range of $\kappa_t^\gamma$ is excluded while all values of $\kappa_t^Z$ remain allowed. These results are not very sensitive to modifications in $z_{cut}$. Clearly, this is *not* the best procedure.

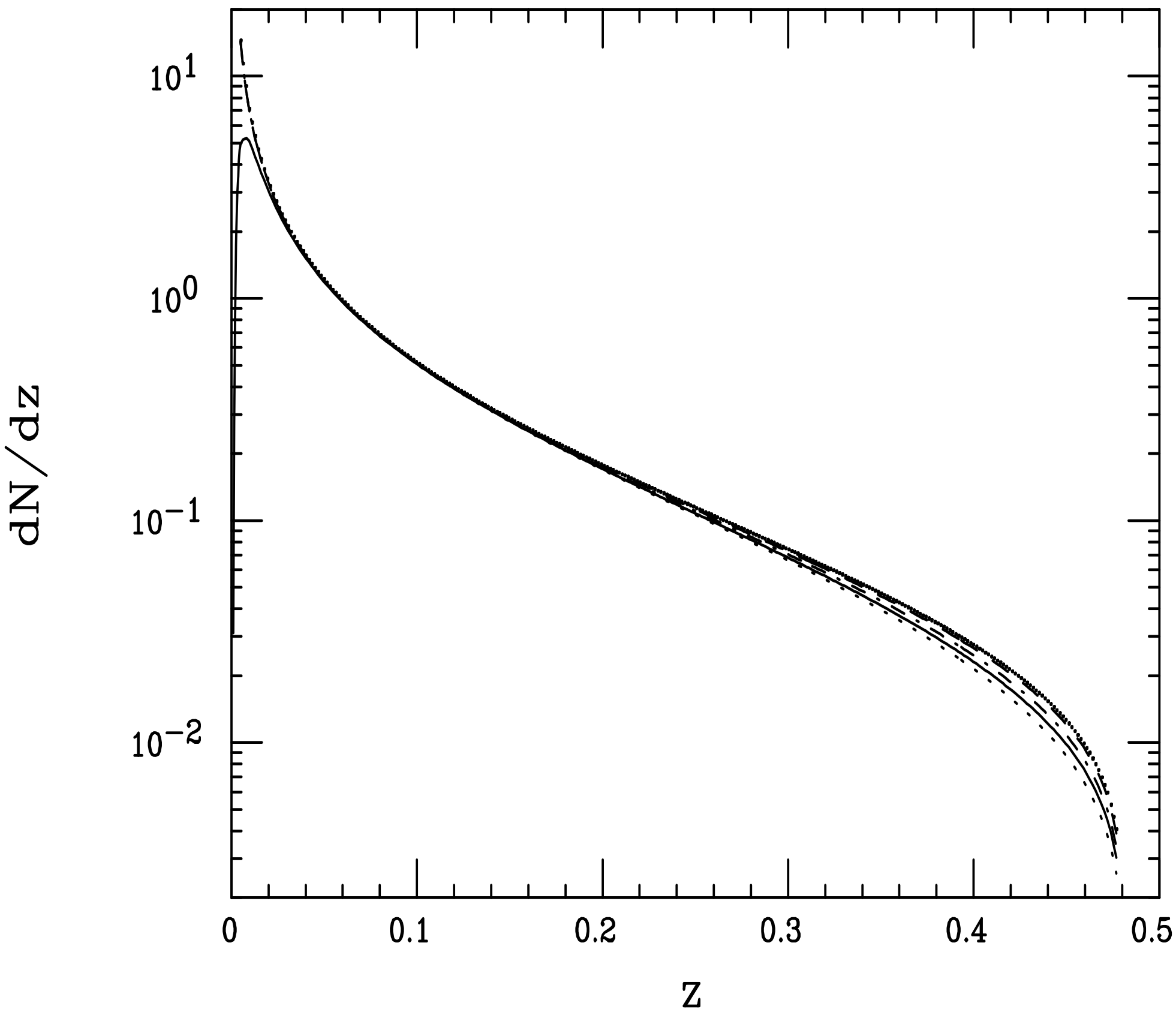

Figure 6: Gluon energy spectrum associated with $t\bar{t}g$ production at a 500 GeV $e^+e^-$ collider assuming $m_t$ = 180 GeV. The solid curve is for the SM and includes the effects of a finite top width. The dotted(dashed, dash-dotted, and square-dotted) curve corresponds to $\kappa_t^\gamma$ =0.2(-0.2) and $\kappa_t^Z$ =0.2(-0.2), respectively.

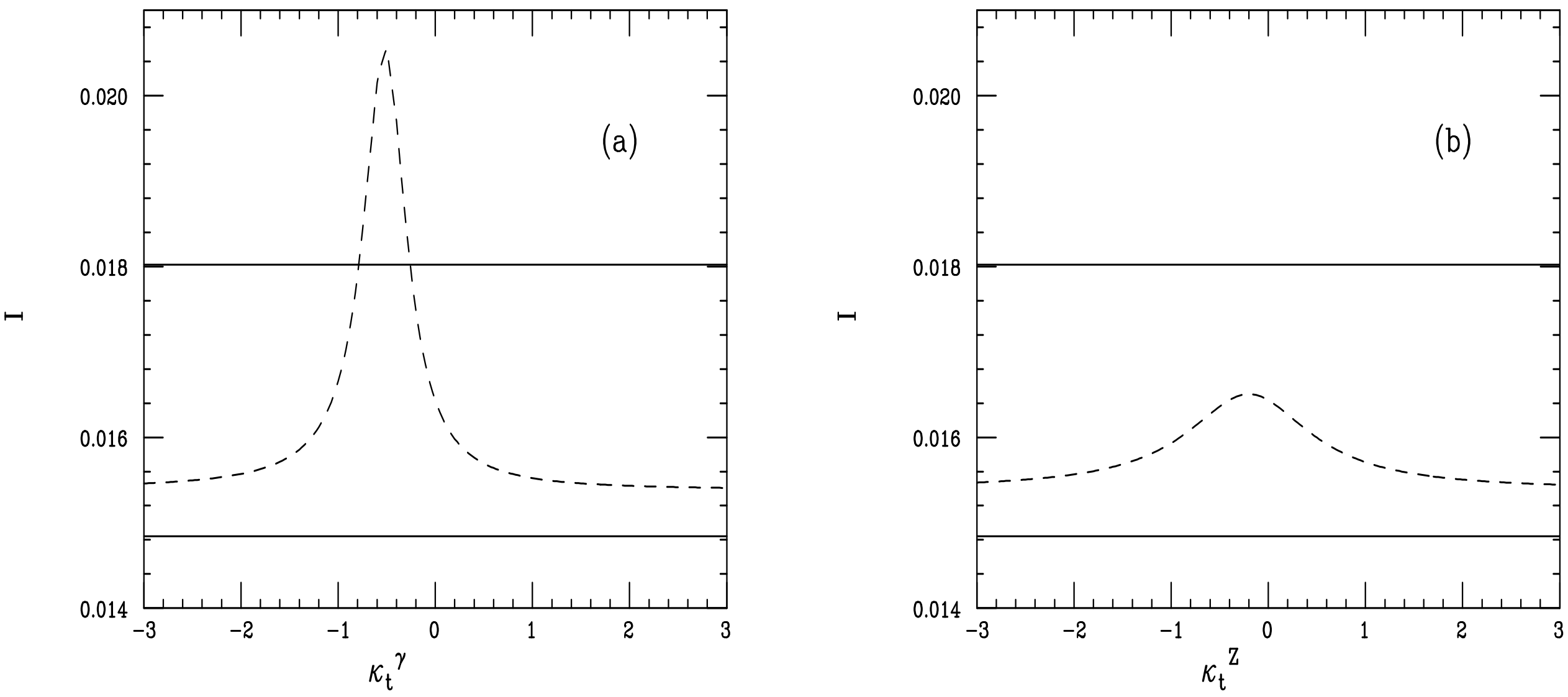

Figure 7: Integrated rate for $z > 0.2$ and $m_t = 180$ GeV as a function of (a) $\kappa_t^\gamma$ or (b) $\kappa_t^Z$ at a 500 GeV $e^+e^-$ collider. The solid lines correspond to the 95% CL bounds accounting for statistical errors only assuming an integrated luminosity of $50 fb^{-1}$.

Next, for purposes of demonstration, we divide the region above $z = 0.15$ into 7 bins of width $\Delta z = 0.05$, except for the highest bin which includes everything above $z = 0.45$. Assuming $\mathcal{L} = 50 fb^{-1}$ and statistical errors only we generate artificial data via a Monte

Carlo assuming the SM is correct and then we fit the resulting distribution allowing for $\kappa_t^\gamma$ or $\kappa_t^Z$ to be non-zero. Allowing $\kappa_t^\gamma$ only to be non-zero yields $\kappa_t^\gamma = 0.009^{+0.027}_{-0.026}$ at 95% CL from the fit. For non-zero $\kappa_t^Z$, we find instead the 95% CL ranges $-0.53 \leq \kappa_t^Z \leq -0.29$ and $-0.09 \leq \kappa_t^Z \leq 0.12$. The two ranges are the result of a double minimum in the $\chi^2$ distribution. Here we see the much greater sensitivity to $\kappa_t^\gamma$ than to $\kappa_t^Z$ as might have been expected from Figs.7a and 7b. The tiny 95% CL range we obtained for $\kappa_t^\gamma$ is clearly an over optimistic result since all systematic errors have been ignored, but it clearly demonstrates that the $\gamma, Zt\bar{t}$ vertices can be probed by using the $t\bar{t}g$ channel. A full Monte Carlo study of this process, including detector effects, would be most enlightening.

The scenario with $\tilde{\kappa}_t^\gamma$ and $\tilde{\kappa}_t^Z$ non-zero is easily analyzed using the previous results by noting that only quadratic terms in these quantities appear in the expressions Eqs.(5-8). In fact, if we average the gluon energy distributions for the cases of positive and negative values of $\kappa_t^\gamma$ we obtain the result for $\tilde{\kappa}_t^\gamma$ and similarly for $\gamma \rightarrow Z$. However, since almost all of the sensitivity to $\kappa_t^{\gamma,Z}$ arises from the linear term in the these equations we will find that the potential constraints on $\tilde{\kappa}_t^{\gamma,Z}$ are relatively weak. From these considerations we obtain, from the Monte Carlo approach described above, that $|\tilde{\kappa}_t^\gamma| \leq 0.296$ and $|\tilde{\kappa}_t^Z| \leq 0.407$ at 95% CL. As in the magnetic weak dipole case we remind the reader that these limits include statistical errors only.

What happens at a higher energy machine? Fig.8 displays the gluon energy spectrum associated with $t\bar{t}g$ production for $m_t = 180$ GeV at a 1 TeV $e^+e^-$ collider for the SM and for the same values of $\kappa_t^\gamma$ and $\kappa_t^Z$ shown in Fig.6. The large $z$ part of this figure indicates that there will be greater sensitivity to the anomalous couplings at these higher energies. This may lead one to re-try our first approach, *i.e.*, just counting the number of events above a fixed value of $z_{cut}$. Figs.9a and 9b show just this situation for $\mathcal{L} = 100 fb^{-1}$ and $z_{cut} = 0.4$, together with the 95% CL bound for the SM used as input assuming only statistical errors as before. Unlike the 500 GeV machine, here we obtain some modest bounds: $-1.1 \leq \kappa_t^\gamma \leq -0.4$ and $-0.1 \leq \kappa_t^\gamma \leq 0.2$ as well as $-0.9 \leq \kappa_t^Z \leq 0.5$. Of course, we expect that by fitting the spectrum we can do even better. We take the region above $z = 0.4$ and divide it into 9 bins of width $\Delta z = 0.05$, except for the last bin as above. Following the same Monte Carlo approach, we obtain $\kappa_t^\gamma = 0.005^{+0.023}_{-0.026}$ and $-0.25 \leq \kappa_t^Z \leq 0.09$, both of which are somewhat stronger than were found above. If we now only assume that the electric weak dipole moments are non-zero, employing the procedure as discussed above for the 500 GeV case yields the corresponding constraints $|\tilde{\kappa}_t^\gamma| \leq 0.118$ and $|\tilde{\kappa}_t^Z| \leq 0.166$ at the 95% CL.

# 4 Summary and Conclusions

In this paper we have considered how the structure of $Z \to b\bar{b}g$ and $e^+e^- \to t\bar{t}g$ events may reveal information on anomalous couplings at the $Zb\bar{b}$ and $\gamma, Zt\bar{t}$ vertices. In the $b$-quark case, two steps were required to perform this analysis:

($i$) The presently allowed ranges of $\kappa_b$ and $\tilde{\kappa}_b$ had to be extracted from the latest round of LEP and SLC data. This required us to update our published analysis using the results presented at Moriond95.

($ii$) The contributions of non-zero $\kappa_b$ and $\tilde{\kappa}_b$ to the differential distributions for $Z \to b\bar{b}g$ had to be determined and scanned over the ranges allowed for these parameters by the electroweak data.

We found that although contributions from possible anomalous weak couplings might have been *a priori* observable in $Z \to b\bar{b}g$, the existing constraints from precision electroweak data are sufficiently tight as to preclude any large effects. Of course, we should continue to probe these couplings by other means.

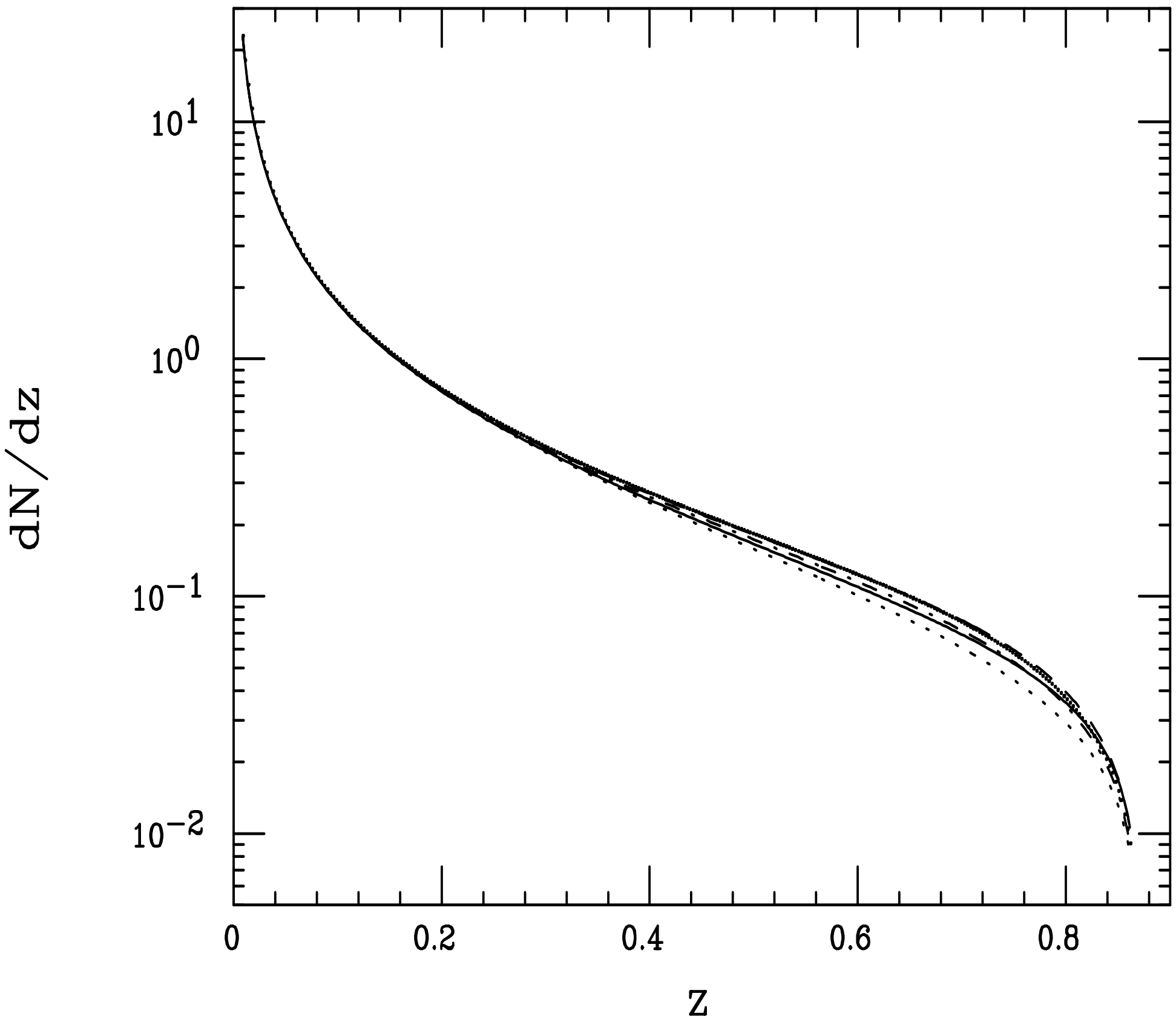

Figure 8: Same as Fig.6 but now for a 1000 GeV $e^+e^-$ collider. Finite top width contributions are ignored.

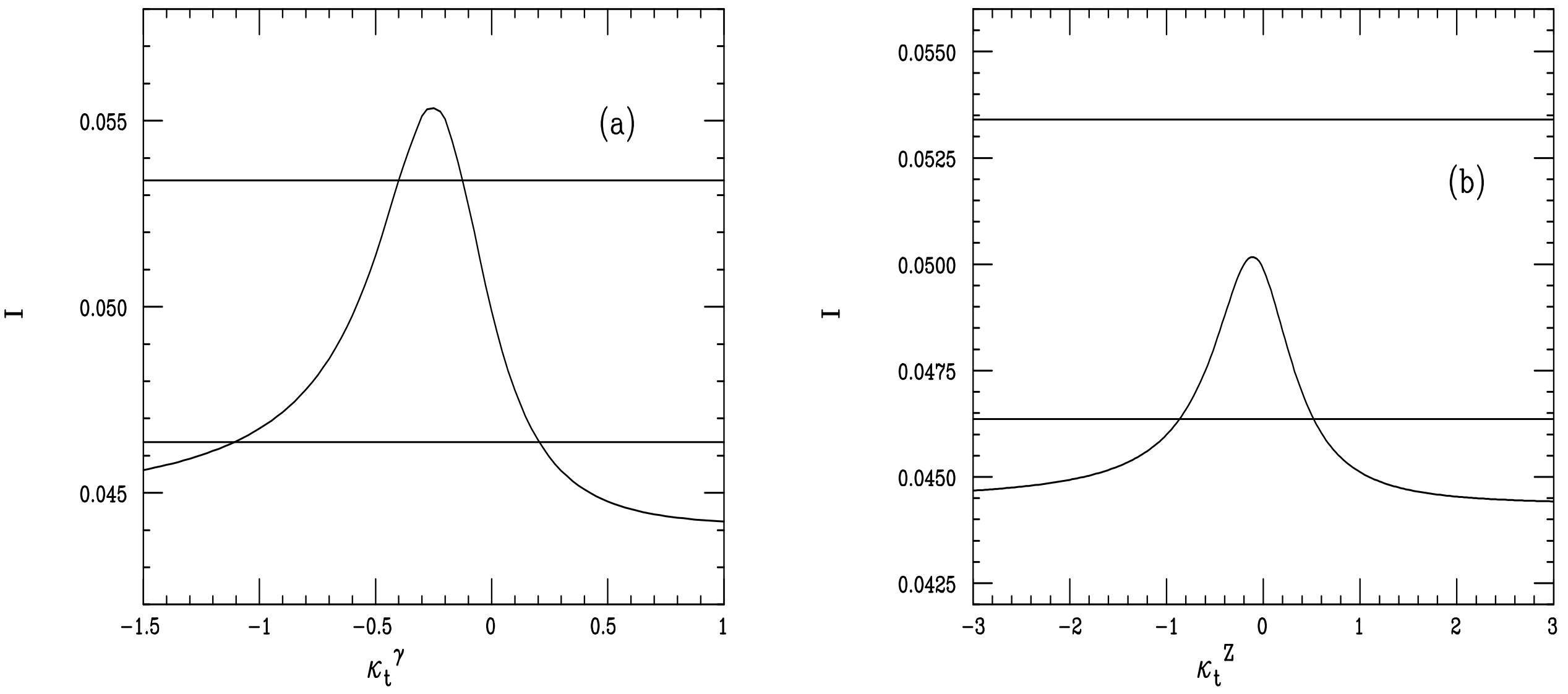

Figure 9: Integrated rate for $z > 0.4$ and $m_t = 180$ GeV as a function of (a) $\kappa_t^\gamma$ or (b) $\kappa_t^Z$ at a 1 TeV $e^+e^-$ collider. The solid lines correspond to the 95% CL bounds accounting for statistical errors only assuming an integrated luminosity of $100 fb^{-1}$.

In the $t$-quark case, we examined the gluon energy distribution, which is the only observable which does not require a detailed analysis of the $t\bar{t}$ decay products. To avoid finite-width effects as well as the contributions due to gluon radiation off of the final state $b$-quarks, we restricted our analysis to gluon energies $\gg \Gamma_t$. Fortunately, this was just the phase space region most sensitive to the $\gamma, Zt\bar{t}$ anomalous couplings we wished to probe. We found that $\kappa_t^\gamma$ is the coupling that we are most sensitive to through fitting the gluon spectrum. The appearance of $\tilde{\kappa}_t^{\gamma,Z}$ only at the quadratic level reduced the sensitivity to their presence, while in the case of $\kappa_t^Z$ a double minimum in the $\chi^2$ distribution also resulted in reduced sensitivity. However, in all cases we found that anomalous couplings are observable with magnitudes comparable to those found through more direct examinations of the $\gamma, Zt\bar{t}$ vertices. In addition, a higher center of mass energy was found to lead to an improvement in the sensitivity to the anomalous couplings. Of course, complete Monte Carlo studies must be performed to determine the true sensitivity to these anomalous couplings and the enhanced capabilities available due to beam polarization must be included.

We should remind the reader before concluding that a deviation in the shape of the spectrum of gluon radiation accompanying $t\bar{t}$ production does not uniquely point to the existence of anomalous $\gamma, Zt\bar{t}$ couplings. As we have shown in our earlier work[20], an analogous modification of the $t\bar{t}g$ coupling can also lead to spectrum shifts. If such deviation are observed experimentally then a detailed analysis will be required to determine the true origin of the effect.

## ACKNOWLEDGEMENTS

The author would like to thank J.L. Hewett, S. Wagner, M. Hildreth and P. Burrows for discussions related to this work. He would also like to thank the members of the Argonne National Laboratory High Energy Theory Group and the Phenomenology Institute at the University of Wisconsin-Madison for their hospitality and use of their computing facilities while this work was in progress.

## APPENDIX

In this Appendix, we provide the exact forms of the expressions used in our analysis. Including both anomalous couplings as well as finite $b$-quark mass effects, the tree level width for $Z \to b\bar{b}$ is given by (again we omit an overall constant which cancels in the ratio)

$$\Gamma(Z \to b\bar{b}) \simeq \frac{3\beta}{4}\Big[(v_b^2 + a_b^2)(1+\beta^2/3) + (v_b^2 - a_b^2)(1-\beta^2)$$

$$+ \frac{r}{4}(1-\beta^2/3)\left[(\kappa_b)^2 + (\tilde{\kappa}_b)^2\right] + (\kappa_b)^2 - (\tilde{\kappa}_b)^2 + 4v_b\kappa_b\Big] ,$$

where $\beta^2 = 1 - 4/r$. In the expressions for the $Z \to b\bar{b}g$ width, the corrections due to finite $r$ are given by the replacements

$$f_{1\kappa} \to f_1 - (z_1 z_2)^{-2}\Big[\frac{2}{r}\left[(z_1^2 + z_2^2) - 6z_1z_2 + 8z_1z_2(z_1+z_2)\right]$$

$$+ \frac{16}{r^2}(z_1+z_2)^2\Big]$$

$$f_{1\tilde{\kappa}} \to f_1 - (z_1 z_2)^{-2}\Big[\frac{2}{r}\left[(z_1^2 + z_2^2) + 6z_1z_2 - 4z_1z_2(z_1+z_2)\right]$$

$$- \frac{8}{r^2}(z_1+z_2)^2\Big]$$

$$f_2 \to f_2 + \frac{12}{r}(z_1+z_2)^2/(z_1z_2)^2 ,$$

Note that the functions for the $\kappa^2$ and $\tilde{\kappa}^2$ terms differ beyond the leading order in $r^{-1}$ so that there are now really two $f_1$ functions. Thus the term $(\kappa^2 + \tilde{\kappa}^2)f_1$ is replaced by $\kappa^2 f_{1\kappa} + \tilde{\kappa}^2 f_{1\tilde{\kappa}}$. As is well known, the usual SM piece is also altered by finite quark mass corrections. Denoting the familiar $(x_1^2+x_2^2)/(1-x_1)(1-x_2)$ expression by $f_0$, we must make the replacement of $(v_b^2 + a_b^2)f_0$ by $v_b^2 f_{0v} + a_b^2 f_{0a}$ where

$$f_{0v} = f_0 + (z_1z_2)^{-2}\Big[\frac{-2}{r}\left[z_1^2(1+2z_2) + z_2^2(1+2z_1)\right]$$

$$
\begin{aligned}
&- \quad \frac{4}{r^2}(z_1+z_2)^2\Big] \\
f_{0a} \quad &= \quad f_0 + (z_1 z_2)^{-2}\left[\frac{-2}{r}\left[-z_1 z_2 (z_1+z_2)^2 - 4 z_1 z_2 (z_1+z_2)\right.\right. \\
&+ \quad \left.\left. z_1^2 + z_2^2 + 6 z_1 z_2\right] + \frac{8}{r^2}(z_1+z_2)^2\right] ,
\end{aligned}
$$

Numerically, as discussed in the text, these higher order terms in $r^{-1}$ are found to be quite small for $b$-quarks in $Z$ decay but would be very important when one looking for the effects of anomalous couplings of the top quark at a high energy $e^+e^-$ collider as discussed in the text.